\documentclass[letterpaper,11pt]{article}
\usepackage[margin=1in]{geometry}
\usepackage{amsfonts,amsmath,amssymb,amsthm,thmtools,mathtools}
\usepackage{booktabs}
\usepackage{enumitem}
\usepackage[dvipsnames]{xcolor}
\usepackage{hyperref}
\usepackage{cleveref}

\hypersetup{
    colorlinks=true,
    linkcolor=RoyalBlue,
    citecolor=ForestGreen,
    urlcolor=RoyalBlue,
    hypertexnames=false
}

\declaretheorem[numberwithin=section]{theorem}
\declaretheorem[sibling=theorem]{lemma}
\declaretheorem[sibling=theorem]{proposition}

\declaretheorem[style=definition,sibling=theorem]{definition}
\declaretheorem[style=definition,sibling=theorem]{remark}

\newcommand{\E}{\mathbb{E}}
\newcommand{\Prob}{\mathbb{P}}
\newcommand{\R}{\mathbb{R}}
\newcommand{\1}{\mathbf{1}}
\newcommand{\ALG}{\mathrm{ALG}}
\newcommand{\Unif}{\mathrm{Unif}}
\newcommand{\Bern}{\mathrm{Bern}}
\newcommand{\TWR}{\mathsf{2S\text{-}RG\text{-}RT}}

\title{Breaking $0.70$ in Prophet Secretary Using Only the Mean:\\
A Simple Two-Stage Running-Max Algorithm and Precommitted Limits}
\author{
Dongchen Li\thanks{School of Computing and Data Science, The University of Hong Kong. Email: \texttt{dongchen.li@connect.hku.hk}.}
}
\date{}

\begin{document}

\begin{titlepage}
\thispagestyle{empty}
\maketitle

\begin{abstract}
In the \emph{Prophet Secretary} problem, $n$ independent nonnegative random variables $X_1, \dots, X_n$ arrive in a uniformly random order, and an online algorithm must irrevocably select at most one item to maximize its expected value relative to the offline optimum $\mu := \E[\max_{1 \le i \le n} X_i] \in (0, \infty)$.
We study the \emph{mean-only} information model, where the algorithm has zero knowledge of the underlying distributions and is given only the single scalar $\mu$ (normalized to $\mu = 1$ by scale invariance).

Our main result breaks the $0.70$ competitive-ratio barrier for Prophet Secretary using a minimal, two-stage adaptive mean-only algorithm $\TWR(\theta, t_1)$ with an explicit 6-segment piecewise-linear threshold $\theta$ and switch time $t_1 = 21/25 = 0.84$:
\begin{itemize}[leftmargin=1.5em]
    \item During Phase~1 ($t < t_1$), an arriving item is accepted iff it exceeds $\theta(t)$ \emph{and} is a strict running maximum (with independent online tie-breakers).
    \item During Phase~2 ($t \ge t_1$), the running-maximum gate is lifted and any item exceeding $\theta(t)$ is accepted.
\end{itemize}
We prove that for \emph{every finite $n \ge 1$} and \emph{every} independent nonnegative distribution with finite positive $\mu$, $\TWR(\theta, t_1)$ achieves competitive ratio strictly exceeding $\mathbf{0.7005} > 0.70$ (Theorem~\ref{thm:main}).
Our proof uses only elementary probabilistic tools: sample-path truncation, an inductive leave-one-out survival monotonicity theorem (Theorem~\ref{thm:loo}), Harris's inequality, and an independent-clock elimination of Phase-2 items (Theorem~\ref{thm:elimB}), followed by a finite count--spatial box cover certified in outward-rounded IEEE-754 interval arithmetic using rigorous right-endpoint lower Riemann sums.

Complementing this positive breakthrough, we show via joint concavity and Jensen's inequality that within the regular precommitted threshold class of Feng, Jin, Wu, and Xiang~\cite{MO26}, even arbitrary $n$-dependent \emph{randomized} regular precommitted thresholds cannot exceed competitive ratio $0.694451$ (Theorem~\ref{thm:rand-ub}), demonstrating that minimal adaptivity separates $\TWR$ from regular precommitted mean-only thresholds.
\end{abstract}
\end{titlepage}

\section{Introduction}

In the \emph{Prophet Secretary} problem~\cite{EHLM17}, $n$ independent nonnegative random variables $X_1, \dots, X_n$ with unknown arrival order are presented in a uniformly random permutation, and an online decision-maker must irrevocably accept at most one item to maximize $\E[\ALG] / \E[X^*]$, where $X^* := \max_{1 \le i \le n} X_i$.
Over the past decade, full-information competitive ratios progressed from $1 - 1/e \approx 0.6321$~\cite{EHLM17,EHKS18}, to $0.669$ via blind quantile strategies~\cite{CSZ21}, and recently to $0.688$~\cite{CHLT25}, while the best known full-information upper bounds stand at $0.7254$~\cite{BC23} and $0.7235$~\cite{GMS23}.
Breaking $0.70$ has remained a central open milestone.

\paragraph{Relationship to prior frameworks.}
Prior approaches use fundamentally different forms of information and adaptivity:
\begin{itemize}[leftmargin=1.5em,itemsep=2pt]
    \item \emph{Blind quantile strategies}~\cite{CSZ21} fix a universal \emph{quantile profile} $\alpha(t)$ (requiring the CDF of $X^*$, but not value ratios) and face a $0.675$ barrier for quantile-blind algorithms.
    \item \emph{Distribution-dependent activation}~\cite{CHLT25} uses full distributional knowledge to design multi-phase activation rules with special treatment of a largest item, reaching $0.688$.
    \item \emph{Mean-only precommitted thresholds}~\cite{MO26} take an orthogonal perspective: the algorithm knows \emph{nothing} about the marginal distributions except the single scalar $\mu = \E[X^*]$, and accepts the first item exceeding a fixed curve $\mu\,\theta(t)$. Because a mean-scaled value threshold is incomparable to a quantile profile, it bypasses the $0.675$ blind-quantile barrier~\cite{CSZ21}, achieving $0.6902$~\cite{MO26}—yet deterministic regular precommitted mean-only thresholds cannot exceed $0.694451$~\cite[Theorem~C.8]{MO26}.
\end{itemize}

\paragraph{Our Contributions.}
Building directly on the mean-only survival and tail-completion perspective of~\cite{MO26}, we show that adding a \emph{single} elementary adaptive ingredient—a Phase-1 running-maximum gate—breaks both the $0.694451$ regular-precommitted barrier and the $0.70$ Prophet Secretary barrier simultaneously:

\begin{enumerate}[leftmargin=2em,itemsep=3pt]
    \item \textbf{Breaking $0.70$ Using Only the Mean (Theorem~\ref{thm:main}).}
    We introduce the two-stage algorithm $\TWR(\theta, t_1)$ with $t_1 = 21/25 = 0.84$ and the explicit 6-segment curve $\theta$ in Table~\ref{tab:curve}. We rigorously prove that for \emph{every finite $n \ge 1$} and \emph{every} product distribution with $\mu = \E[X^*] \in (0, \infty)$,
    \[
        \E[\ALG] \;>\; \mathbf{0.7005} \cdot \E[X^*].
    \]
    The running-maximum gate provides a clean, distribution-free form of adaptivity: rather than identifying a largest item from known distributions as in~\cite{CHLT25}, it uses online record status to neutralize high-density floors while never rejecting the global maximum.

    \item \textbf{Minimal, Elementary Proof Architecture.}
    The entire reduction uses only four elementary ingredients:
    \begin{itemize}[leftmargin=1.5em,itemsep=1pt]
        \item \emph{Truncation invariance and exact tail identity (Lemma~\ref{lem:trunc}):} Capping all items at $\theta(0)$ leaves the acceptance time invariant and yields $\Prob(X_\sigma > x) = \sum_{i=1}^n \Prob(X_i > x)\E[\tau^{-i}]$ for all $x > \theta(0)$.
        \item \emph{Leave-one-out survival monotonicity (Theorem~\ref{thm:loo}):} Even though an early-arriving large item can block smaller items during Phase~1, deleting any item \emph{always} weakly increases the survival probability $S(t; u)$ at every $t \in [0, 1]$. Combined with coordinate-wise monotonicity (Lemma~\ref{lem:coord-mono}) and Harris's inequality (Lemma~\ref{lem:harris}), this yields a lossless finite-$n$ \emph{Master Reduction} (Theorem~\ref{thm:master}) to a deterministic functional $G(u)$.
        \item \emph{Independent-clock Phase-2 elimination (Theorem~\ref{thm:elimB}):} Conditioning on the first eligible Phase-2 arrival time $Z$ of items with $u_j \ge t_1$—which is independent of the Phase-1 item process—eliminates \emph{all} Phase-2 items in three lines with no derivatives or Fubini bookkeeping!
        \item \emph{Count--spatial box cover with outward-rounded interval verification (Section~\ref{sec:cert}):} Applying Theorem~\ref{thm:loo} and Lemma~\ref{lem:coord-mono} once more gives explicit lower- and upper-dominating configurations $(A^-, A^+)$ for each count--spatial box, whose lower bound is certified via rigorous one-sided right-endpoint Riemann sums evaluated in outward-rounded IEEE-754 interval arithmetic over exact rational grids.
    \end{itemize}

    \item \textbf{Randomized Precommitted Impossibility (Theorem~\ref{thm:rand-ub}).}
    In Section~\ref{sec:precommitted-ub}, we show that for arbitrary $n$-dependent \emph{randomized} distributions $\Theta_n$ over the regular precommitted threshold class of~\cite[Theorem~C.8]{MO26} (continuous, strictly decreasing $\theta : [0, 1] \to [0, 1]$ with $\theta(1) = 0$), joint concavity of the limiting reward and Jensen's inequality imply an asymptotic upper bound of $0.694451$. Thus a sharp separation holds between regular precommitted mean-only thresholds ($\le 0.694451$) and minimally adaptive mean-only thresholds ($> 0.7005$).
\end{enumerate}

\section{Model and the Two-Stage Algorithm \texorpdfstring{$\TWR(\theta, t_1)$}{2S-RG-RT}}
\label{sec:model}

\subsection{Continuous-Time Prophet Secretary and Mean-Only Setting}

Let $X_1, \dots, X_n$ ($n \ge 1$) be independent nonnegative random variables with arbitrary marginal distributions $F_1, \dots, F_n$, and let $X^* := \max_{1 \le i \le n} X_i$ with finite positive mean $\mu := \E[X^*] \in (0, \infty)$.
Because our algorithm scales its threshold linearly in $\mu$, we normalize $\mu = \E[X^*] = 1$ without loss of generality.

We work in the standard continuous-time uniform-arrival model~\cite{EHKS18,CSZ21}: each item $i \in [n] := \{1, \dots, n\}$ draws an independent arrival time $T_i \sim \Unif[0, 1]$, independent of $X = (X_1, \dots, X_n)$, and items arrive in increasing order of $T_i$.
For every finite $n \ge 1$, this is strictly equivalent to the discrete uniformly random arrival-order model: given $n$ (or directly in the random-arrival-time model without knowing $n$), an online algorithm samples $n$ i.i.d.\ $\Unif[0,1]$ values upfront, sorts them as $0 < T_{(1)} < \dots < T_{(n)} < 1$, and assigns $T_{(k)}$ to the $k$-th arriving item.
All positive results in this paper hold for \emph{every finite $n \ge 1$}.

\subsection{Algorithm \texorpdfstring{$\TWR(\theta, t_1)$}{2S-RG-RT}}

Fix a continuous, strictly decreasing threshold function $\theta : [0, 1] \to [0, \theta(0)]$ with $\theta(1) = 0$, and a phase-switch time $t_1 \in (0, 1)$.
Let $\succ$ denote the strict lexicographic order on $\R^2$:
\[
    (x, u) \succ (x', u') \iff x > x' \;\;\text{or}\;\; (x = x' \text{ and } u > u').
\]

\begin{definition}[Algorithm $\TWR(\theta, t_1)$]
Initialize $(M_X, M_U) \leftarrow (-1, -1)$. When item $i$ arrives at time $T_i \in [0, 1]$ with value $X_i \ge 0$ (and no item has been accepted yet):
\begin{enumerate}[leftmargin=2em,itemsep=2pt]
    \item Draw an independent online tie-breaker $U_i \sim \Unif[0, 1]$.
    \item Set $\mathsf{rmax}_i := \1\{(X_i, U_i) \succ (M_X, M_U)\}$.
    \item \textbf{Phase~1 ($T_i < t_1$):} Accept item $i$ and stop iff $X_i > \theta(T_i)$ \emph{and} $\mathsf{rmax}_i = 1$.
    \item \textbf{Phase~2 ($T_i \ge t_1$):} Accept item $i$ and stop iff $X_i > \theta(T_i)$.
    \item If item $i$ is not accepted and $\mathsf{rmax}_i = 1$, update $(M_X, M_U) \leftarrow (X_i, U_i)$.
\end{enumerate}
Define the \emph{acceptance time} $\sigma := T_i$ if item $i$ is accepted, and $\sigma := \infty$ (with reward $X_\sigma := 0$) if no item is accepted.
Define the bounded \emph{stopping time} $\tau := \min(\sigma, 1) \in [0, 1]$, and the \emph{survival probability}
\[
    S(t) \;:=\; \Prob(\sigma > t), \qquad t \in [0, 1].
\]
Thus $S(1) = \Prob(\sigma = \infty)$ is the exact probability of never accepting any item, and $\E[\tau] = \int_0^1 S(t)\, dt$.
\end{definition}

For Theorem~\ref{thm:main}, we fix the exact rational switch time $t_1 = 21/25 = 0.84$ and the continuous 6-segment piecewise-linear curve $\theta : [0, 1] \to [0, 37/50]$ linearly interpolating the 7 exact rational knots in Table~\ref{tab:curve}.

\begin{table}[ht]
\centering
\caption{Exact rational knots of the 6-segment threshold $\theta(t)$ for $\TWR(\theta, t_1 = 21/25)$.}
\label{tab:curve}
\begin{tabular}{cccccccc}
\toprule
$t$ & $0$ & $1/5\ (0.20)$ & $2/5\ (0.40)$ & $3/5\ (0.60)$ & $21/25\ (0.84)$ & $23/25\ (0.92)$ & $1$ \\
\midrule
$\theta(t)$ & $37/50\ (0.740)$ & $119/200\ (0.595)$ & $23/50\ (0.460)$ & $69/200\ (0.345)$ & $8/25\ (0.320)$ & $29/200\ (0.145)$ & $0$ \\
\bottomrule
\end{tabular}
\end{table}

\paragraph{Intuition.}
High-density deterministic floors ($k$ identical copies of $v < \theta(0)$ with eligibility time $u = \theta^{-1}(v)$) collapse a precommitted threshold's stopping time to $u$ as $k \to \infty$, sacrificing rare high-value items. Under $\TWR$, by Lemma~\ref{lem:surv} the Phase-1 survival of $k$ identical copies is $S_k(t) = \frac{u}{t} + \big(1 - \frac{u}{t}\big)(1 - t)^k \to \frac{u}{t}$ as $k \to \infty$ (since the highest-tie-breaker copy arriving before $t$ blocks later arrivals whenever it arrives before $u$), raising limiting survival from $0$ to $u/t$ on $[u, t_1]$. Lifting the running-max gate at $t_1 = 0.84$ ensures that an early-arriving large item does not permanently block recovery of secondary items on $[t_1, 1]$, while independent tie-breakers $U_i$ prevent identical-item blocking degeneracy.

\section{Survival Identities, Monotonicity, and Deterministic Master Reduction}
\label{sec:master}

Let $\widetilde X_i := \min(X_i, \theta(0))$ and $\widetilde X^* := \max_{i \in [n]} \widetilde X_i = \min(X^*, \theta(0))$.
For each $i \in [n]$, let $\sigma^{-i}$ and $\tau^{-i} := \min(\sigma^{-i}, 1)$ denote the acceptance and stopping times of $\TWR(\theta, t_1)$ run on the $(n-1)$-item instance $(X_{-i}, T_{-i}, U_{-i})$ obtained by deleting item $i$ (for $n = 1$, the empty instance has $\sigma^{-1} = \infty$ and $\tau^{-1} = 1$).

\subsection{Truncation Invariance and Exact Tail Identity}

\begin{lemma}[Truncation Invariance and Exact Tail Identity]
\label{lem:trunc}
For every finite $n \ge 1$ and every independent nonnegative $(X_1, \dots, X_n)$:
\begin{enumerate}[leftmargin=2em,itemsep=2pt]
    \item \textbf{Sample-path invariance:} Almost surely over $(T, U)$, running $\TWR(\theta, t_1)$ on $(X, T, U)$ and on $(\widetilde X, T, U)$ yields the exact same acceptance time $\sigma(X, T, U) = \sigma(\widetilde X, T, U)$ (hence $\tau(X, T, U) = \tau(\widetilde X, T, U)$) and satisfies $\widetilde X_{\sigma(X)} = \widetilde X_{\sigma(\widetilde X)}$.
    \item \textbf{Exact tail identity:} For every $x > \theta(0)$,
    \[
        \Prob(X_\sigma > x) \;=\; \sum_{i=1}^n \Prob(X_i > x)\,\E[\tau^{-i}].
    \]
\end{enumerate}
\end{lemma}

\begin{proof}
Work on the full-probability event where all $T_i \in (0, 1)$ are distinct and all $(X_i, U_i)$ are distinct.

\emph{Part 1.} Let $H := \{i \in [n] : X_i \ge \theta(0)\}$. If $H = \emptyset$, then $X = \widetilde X$ and the claim is immediate.
Otherwise, let $i_0 := \arg\min_{i \in H} T_i$. Every item $j$ arriving before $T_{i_0}$ has $X_j < \theta(0)$, so $X_j = \widetilde X_j$; thus both executions agree identically on $[0, T_{i_0})$.
If neither has accepted before $T_{i_0}$, then at time $T_{i_0} > 0$, strict decrease of $\theta$ gives $X_{i_0} \ge \widetilde X_{i_0} = \theta(0) > \theta(T_{i_0})$, and every earlier item $j$ satisfies $X_j = \widetilde X_j < \theta(0) = \widetilde X_{i_0} \le X_{i_0}$.
Hence $i_0$ exceeds $\theta(T_{i_0})$ and is strictly the running maximum in both executions, so both accept $i_0$ at time $T_{i_0}$.

\emph{Part 2.} Fix $x > \theta(0)$ and $i \in [n]$, and suppose $X_i > x > \theta(0)$.
If $\tau^{-i} < T_i$ (equivalently $\sigma^{-i} < T_i$), the full instance already accepted an item before $T_i$, so $i$ is not accepted.
Conversely, if $\tau^{-i} > T_i$ (so $\sigma^{-i} > T_i$), no item was accepted before $T_i$, and no $j \ne i$ with $X_j \ge \theta(0)$ can have arrived before $T_i$ (else, by Part~1, the earliest such item in $X_{-i}$ would have been accepted before $T_i$, contradicting $\sigma^{-i} > T_i$).
Hence every item arriving before $T_i$ has value $< \theta(0) < X_i$, so $i$ exceeds $\theta(T_i)$ and is strictly the running maximum at $T_i$, and is accepted.
Since $\Prob(T_i = \tau^{-i}) = 0$, we have $\1\{\text{accept } i \text{ and } X_i > x\} = \1\{X_i > x\}\,\1\{T_i < \tau^{-i}\}$ almost surely.
By mutual independence of $X_i$, $T_i \sim \Unif[0, 1]$, and $(X_{-i}, T_{-i}, U_{-i})$, taking expectations and summing over $i \in [n]$ gives $\Prob(X_\sigma > x) = \sum_{i=1}^n \Prob(X_i > x)\,\Prob(T_i < \tau^{-i}) = \sum_{i=1}^n \Prob(X_i > x)\,\E[\tau^{-i}]$.
\end{proof}

\subsection{Exact Survival Formulas}

Fix any deterministic truncated realization $\widetilde X = (x_1, \dots, x_n) \in [0, \theta(0)]^n$.
Condition on the realized tie-breakers $U_1, \dots, U_n$ (almost surely all $(x_i, U_i)$ are distinct), and relabel the items in \emph{strictly decreasing} lexicographic order: $(x_1, U_1) \succ (x_2, U_2) \succ \dots \succ (x_n, U_n)$.
Define their \emph{eligibility times}
\[
    u_j \;:=\; \theta^{-1}(x_j) \in [0, 1], \qquad 0 \le u_1 \le u_2 \le \dots \le u_n \le 1,
\]
where $\theta^{-1}(\theta(0)) = 0$ and $\theta^{-1}(0) = 1$.
Let $S(t; u_1, \dots, u_n) := \Prob_T(\sigma > t \mid u_1, \dots, u_n)$ for $t \in [0, 1]$, with $S(t; \emptyset) \equiv 1$.

\begin{lemma}[Exact Survival Formulas]
\label{lem:surv}
For every $n \ge 1$ and $0 \le u_1 \le \dots \le u_n \le 1$, the survival probability $S(t; u)$ depends only on $u$ (independent of the realized values of $U_i$) and satisfies:
\begin{enumerate}[leftmargin=2em,itemsep=2pt]
    \item \textbf{Phase~1 ($t \in [0, t_1]$):} Let $k := \#\{j \in [n] : u_j < t\}$. If $k = 0$, then $S(t; u) = 1$; if $k \ge 1$, then
    \begin{equation}
    \label{eq:S-p1}
        S(t; u_1, \dots, u_n) = u_1 + (1 - t)\, S(t; u_2, \dots, u_n) = \sum_{j=1}^k (1 - t)^{j-1} u_j + (1 - t)^k.
    \end{equation}
    \item \textbf{Phase~2 ($t \in (t_1, 1]$):} Define $(z)_+ := \max(z, 0)$ and $P(t; v_1, \dots, v_m) := \prod_{r=1}^m (1 - (t - \max(t_1, v_r))_+)$.
    \begin{itemize}[leftmargin=1.5em]
        \item If $u_1 \ge t_1$, then $S(t; u_1, \dots, u_n) = \prod_{j=1}^n \big(1 - (t - u_j)_+\big) = P(t; u_1, \dots, u_n)$.
        \item If $u_1 < t_1 < t$, then
        \begin{equation}
        \label{eq:S-p2}
            S(t; u_1, \dots, u_n) = u_1\, P(t; u_2, \dots, u_n) + (1 - t)\, S(t; u_2, \dots, u_n).
        \end{equation}
    \end{itemize}
\end{enumerate}
\end{lemma}

\begin{proof}
Condition on $T_1 \sim \Unif[0, 1]$, independent of $T_2, \dots, T_n$.

\emph{Phase~1 ($u_1 < t \le t_1$).}
(i) If $T_1 \in [0, u_1)$ (prob.\ $u_1$): no item is eligible on $[0, T_1]$ since $u_j \ge u_1 > T_1$ for all $j \ge 2$, and on $(T_1, t] \subset (T_1, t_1]$ item~$1$ strictly dominates all other items in $(x_j, U_j)$ and blocks every subsequent item's running-max condition; conditional probability of $\sigma > t$ is $1$.
(ii) If $T_1 \in (u_1, t)$ (prob.\ $t - u_1$): if not already stopped before $T_1$, item~$1$ is both eligible and the strict running maximum at $T_1$, so $\sigma \le T_1 < t$; conditional survival is $0$.
(iii) If $T_1 \in (t, 1]$ (prob.\ $1 - t$): item~$1$ has not arrived by $t$, so conditional survival is $S(t; u_2, \dots, u_n)$.
Summing yields~\eqref{eq:S-p1}, and iterating $k$ times gives the closed form.

\emph{Phase~2 ($t \in (t_1, 1]$).}
If $u_1 \ge t_1$, no item is eligible in Phase~1, and in Phase~2 the running-max gate is off, so $\sigma > t$ iff $T_j \notin (u_j, t)$ for all $j \in [n]$ (note $u_j = 1$ gives factor $1$).
If $u_1 < t_1 < t$: when $T_1 \in [0, u_1)$ (prob.\ $u_1$), the algorithm survives Phase~1 with conditional probability $1$ and then survives $(t_1, t]$ (gate off) iff each $j \ge 2$ avoids $(\max(t_1, u_j), t)$, which has probability $P(t; u_2, \dots, u_n)$; when $T_1 \in (u_1, t)$, conditional survival is $0$; when $T_1 > t$ (prob.\ $1 - t$), conditional survival is $S(t; u_2, \dots, u_n)$. Combining gives~\eqref{eq:S-p2} (including at $t = 1$, where $1 - t = 0$ and $S(1; u) = u_1 P(1; u_{-1})$).
\end{proof}

\subsection{Leave-One-Out and Coordinate-Wise Monotonicity}

Let $u_{-j}$ denote $u = (u_1, \dots, u_n)$ with its $j$-th coordinate deleted.

\begin{theorem}[Universal Leave-One-Out Survival Monotonicity]
\label{thm:loo}
For every $n \ge 1$, every $0 \le u_1 \le \dots \le u_n \le 1$, every $j \in [n]$, and every $t \in [0, 1]$,
\[
    S(t; u_{-j}) \;\ge\; S(t; u).
\]
In particular, for every product distribution and every $i \in [n]$, $\E[\tau^{-i}] \ge \E[\tau]$.
\end{theorem}

\begin{proof}
Extend $P(t; v) \equiv 1$ for $t \le t_1$. We first prove by induction on $m \ge 0$ that for every $0 \le \alpha \le v_1 \le \dots \le v_m \le 1$ and every $t \in (\alpha, 1]$,
\begin{equation}
\label{eq:aux-star}
    S(t; v_1, \dots, v_m) \;\ge\; \frac{\alpha}{t}\, P(t; v_1, \dots, v_m).
\end{equation}
For $m = 0$, $1 \ge \alpha/t$ is trivial (and for $\alpha = 0$, the right-hand side is $0$). For $m \ge 1$:
\begin{itemize}[leftmargin=1.5em,itemsep=1pt]
    \item If $t \le t_1$ and $v_1 \ge t$, then $S(t; v) = 1 \ge \alpha/t$. If $v_1 < t \le t_1$, the induction hypothesis gives $S(t; v_{-1}) \ge \alpha/t$, so $S(t; v) = v_1 + (1-t)S(t; v_{-1}) \ge \alpha + (1-t)\frac{\alpha}{t} = \frac{\alpha}{t}$.
    \item If $t > t_1$ and $v_1 \ge t_1$, then $S(t; v) = P(t; v) \ge \frac{\alpha}{t} P(t; v)$. If $v_1 < t_1 < t$, then $P(t; v) = (1 - t + t_1) P(t; v_{-1}) \le P(t; v_{-1})$ while induction gives $S(t; v_{-1}) \ge \frac{\alpha}{t} P(t; v_{-1})$, so
    \[
        S(t; v) = v_1 P(t; v_{-1}) + (1-t) S(t; v_{-1}) \ge \Big(\alpha + (1-t)\frac{\alpha}{t}\Big) P(t; v_{-1}) = \frac{\alpha}{t} P(t; v_{-1}) \ge \frac{\alpha}{t} P(t; v).
    \]
\end{itemize}
Now prove $S(t; u_{-j}) \ge S(t; u)$ by induction on $n$. For $n = 1$, $S(t; \emptyset) = 1 \ge S(t; u_1)$. For $n \ge 2$:
\begin{itemize}[leftmargin=1.5em,itemsep=1pt]
    \item \emph{Case $t \le t_1$:} If $u_1 \ge t$, then $S(t; u) = 1 = S(t; u_{-j})$. Suppose $u_1 < t$.
    If $j = 1$, applying~\eqref{eq:aux-star} with $\alpha = u_1 < t$ and $v = u_{-1}$ gives $u_1 \le t\, S(t; u_{-1})$, so $S(t; u) = u_1 + (1-t) S(t; u_{-1}) \le S(t; u_{-1})$.
    If $j \ge 2$, the induction hypothesis gives $S(t; u_{-1}) \le S(t; (u_{-1})_{-j})$, hence $S(t; u) \le S(t; u_{-j})$.
    \item \emph{Case $t > t_1$:} If $u_1 \ge t_1$, deleting a factor $(1 - (t - u_j)_+) \in [0, 1]$ from $P(t; u)$ cannot decrease the product. Suppose $u_1 < t_1$.
    If $j = 1$, applying~\eqref{eq:aux-star} with $\alpha = u_1 < t_1 < t$ and $v = u_{-1}$ gives $u_1 P(t; u_{-1}) \le t\, S(t; u_{-1})$, so $S(t; u) = u_1 P(t; u_{-1}) + (1-t) S(t; u_{-1}) \le S(t; u_{-1})$.
    If $j \ge 2$, both $P(t; u_{-1}) \le P(t; (u_{-1})_{-j})$ and (by induction) $S(t; u_{-1}) \le S(t; (u_{-1})_{-j})$, giving $S(t; u) \le S(t; u_{-j})$.
\end{itemize}
Integrating over $t \in [0, 1]$ and taking expectation over $\widetilde X$ yields $\E[\tau^{-i}] = \E[\int_0^1 S(t; u_{-i})\, dt] \ge \E[\int_0^1 S(t; u)\, dt] = \E[\tau]$.
\end{proof}

\begin{lemma}[Continuity and Coordinate-Wise Monotonicity]
\label{lem:coord-mono}
For every fixed $t \in [0, 1]$, viewed as a symmetric function of $u = (u_1, \dots, u_n) \in [0, 1]^n$, $S(t; u)$ is continuous and coordinate-wise nondecreasing.
Consequently, $\E_{T,U}[\tau \mid \widetilde X] = \int_0^1 S(t; \theta^{-1}(\widetilde X_1), \dots, \theta^{-1}(\widetilde X_n))\, dt$ is coordinate-wise nonincreasing in $(\widetilde X_1, \dots, \widetilde X_n)$.
\end{lemma}

\begin{proof}
By symmetry, it suffices to verify continuity on $0 \le u_1 \le \dots \le u_n \le 1$ and $\partial S(t; u)/\partial u_j \ge 0$ on $u_1 < \dots < u_n$.
For $t \le t_1$, \eqref{eq:S-p1} has $\partial S/\partial u_j = (1-t)^{j-1} > 0$ for $u_j < t$ and $0$ for $u_j > t$, and glues continuously at $u_j = t$ since $(1-t)^{j-1} t + (1-t)^j = (1-t)^{j-1}$.
For $t > t_1$, expanding~\eqref{eq:S-p2} for $u_1 \le \dots \le u_k < t_1 \le u_{k+1} \le \dots \le u_n$ gives
\begin{equation}
\label{eq:S-p2-closed}
    S(t; u) = \left[\sum_{j=1}^k (1-t)^{j-1} u_j (1 - t + t_1)^{k-j} + (1-t)^k\right] \prod_{r=k+1}^n \big(1 - (t - u_r)_+\big).
\end{equation}
This is positively linear in each $u_j < t_1$, nondecreasing and continuous in each $u_r > t_1$, and continuous as $u_k \to t_1^-$ since $(1-t)^{k-1} t_1 + (1-t)^k = (1-t)^{k-1}(1 - t + t_1)$. Since $\theta^{-1}$ is strictly decreasing, $\E_{T,U}[\tau \mid \widetilde X]$ is coordinate-wise nonincreasing in $\widetilde X$.
\end{proof}

\subsection{Harris's Inequality and the Deterministic Master Reduction}

\begin{lemma}[Harris's Inequality]
\label{lem:harris}
Let $Z_1, \dots, Z_n$ be independent real-valued random variables, and let $f, g : \R^n \to \R$ be bounded measurable functions with $f$ coordinate-wise nondecreasing and $g$ coordinate-wise nonincreasing. Then $\E[f(Z) g(Z)] \le \E[f(Z)]\,\E[g(Z)]$.
\end{lemma}

\begin{proof}
Induct on $n$. For $n = 1$, let $Z_1'$ be an i.i.d.\ copy of $Z_1$; then $(f(Z_1) - f(Z_1'))(g(Z_1) - g(Z_1')) \le 0$, and taking expectations yields $\E[f(Z_1)g(Z_1)] \le \E[f(Z_1)]\E[g(Z_1)]$.
For $n \ge 2$, let $\bar f(z) := \E[f(Z_1, \dots, Z_{n-1}, z)]$ (nondecreasing) and $\bar g(z) := \E[g(Z_1, \dots, Z_{n-1}, z)]$ (nonincreasing). Conditioning on $Z_n$ and applying the induction hypothesis followed by the $n = 1$ case gives
\[
    \E[f(Z) g(Z)] = \E_{Z_n}\!\big[\E[f(Z) g(Z) \mid Z_n]\big] \le \E_{Z_n}\!\big[\bar f(Z_n) \bar g(Z_n)\big] \le \E[\bar f(Z_n)]\,\E[\bar g(Z_n)] = \E[f(Z)]\,\E[g(Z)]. \qedhere
\]
\end{proof}

For any finite sorted eligibility vector $u = (u_1 \le \dots \le u_m) \in [0, 1]^m$ (representing deterministic items $x_j = \theta(u_j)$), define the deterministic functional
\begin{equation}
\label{eq:G-def}
    G(u_1, \dots, u_m) \;:=\; \E_{T,U}\big[\ALG(u)\big] \;+\; \big(1 - \theta(u_1)\big)\,\E_{T,U}\big[\tau(u)\big].
\end{equation}

\begin{theorem}[Deterministic Master Reduction]
\label{thm:master}
For every finite $n \ge 1$ and every independent nonnegative distribution $(X_1, \dots, X_n)$ with $\E[X^*] = 1$,
\[
    \E[X_\sigma] \;\ge\; \E_X\big[G(\theta^{-1}(\widetilde X))\big] \;\ge\; \inf_{m \ge 1}\;\inf_{0 \le u_1 \le \dots \le u_m \le 1} G(u_1, \dots, u_m).
\]
\end{theorem}

\begin{proof}
Decompose $\E[X_\sigma] = \E[\widetilde X_\sigma] + \int_{\theta(0)}^\infty \Prob(X_\sigma > x)\, dx$.
Applying Lemma~\ref{lem:trunc}(2), Theorem~\ref{thm:loo} ($\E[\tau^{-i}] \ge \E[\tau]$), Tonelli's theorem, and $\sum_{i=1}^n \Prob(X_i > x) \ge \Prob(X^* > x)$ gives
\[
    \int_{\theta(0)}^\infty \Prob(X_\sigma > x)\, dx
    = \int_{\theta(0)}^\infty \sum_{i=1}^n \Prob(X_i > x)\,\E[\tau^{-i}]\, dx
    \ge \E[\tau] \int_{\theta(0)}^\infty \Prob(X^* > x)\, dx
    = (1 - \E[\widetilde X^*])\,\E[\tau].
\]
By Lemma~\ref{lem:coord-mono} and Lemma~\ref{lem:harris} applied to the independent variables $\widetilde X_1, \dots, \widetilde X_n$,
\[
    (1 - \E[\widetilde X^*])\,\E[\tau] \;\ge\; \E_X\big[(1 - \widetilde X^*)\,\E_{T,U}[\tau \mid \widetilde X]\big].
\]
Adding $\E[\widetilde X_\sigma] = \E_X[\E_{T,U}[\ALG(\widetilde X) \mid \widetilde X]]$ (Lemma~\ref{lem:trunc}(1)) completes the proof.
\end{proof}

\section{Independent-Clock Elimination of Phase-2 Items}
\label{sec:elimB}

Fix any finite sorted eligibility vector $u = (u_1 \le \dots \le u_m) \in [0, 1]^m$, and split it at $t_1$ into
\[
    A := \{j \in [m] : u_j < t_1\} = (u_1, \dots, u_k), \qquad B := \{j \in [m] : u_j \ge t_1\} = (u_{k+1}, \dots, u_m).
\]
We eliminate all Phase-2 items $B$ in three lines by conditioning on a single independent random clock.

\begin{theorem}[Independent-Clock Phase-2 Elimination]
\label{thm:elimB}
\leavevmode
\begin{enumerate}[leftmargin=2em,itemsep=3pt]
    \item \textbf{Case $A \ne \emptyset$ ($k \ge 1$):} For every $B \subset [t_1, 1]$,
    \[
        G(A \cup B) \;\ge\; \min_{s^* \in [t_1, 1]} G_{\mathrm{elimB}}(A; s^*),
    \]
    where for every $s^* \in [t_1, 1]$,
    \begin{equation}
    \label{eq:G-elimB}
        G_{\mathrm{elimB}}(A; s^*) \;:=\; \theta(t_1) - \big(\theta(t_1) - \theta(s^*)\big) S_A(s^*)
        + \big(1 - \theta(u_1)\big) \int_0^{s^*} S_A(t)\, dt
        + \sum_{j=1}^k \big(\theta(u_j) - \theta(t_1)\big)\big(\pi_j^{\mathrm{P1}}(A) + \pi_j^{\mathrm{P2}}(A; s^*)\big),
    \end{equation}
    with $\pi_j^{\mathrm{P1}}(A) := \int_{u_j}^{t_1} (1 - t)^{j-1} S_{A, >j}(t)\, dt$ and $\pi_j^{\mathrm{P2}}(A; s^*) := \int_{t_1}^{s^*} S_{A, -j}(t)\, dt$ (here $A_{>j} := (u_{j+1}, \dots, u_k)$).

    \item \textbf{Case $A = \emptyset$ ($t_1 \le u_1 \le \dots \le u_m \le 1$):}
    \[
        G(\emptyset \cup B) \;\ge\; \min_{t_1 \le u_1 \le s^* \le 1} G_{\mathrm{P2}}(u_1, s^*),
    \]
    where
    \begin{equation}
    \label{eq:G-P2}
        G_{\mathrm{P2}}(u_1, s^*) \;:=\; \theta(u_1)(s^* - u_1) + \theta(s^*)\big(1 - (s^* - u_1)\big) + \big(1 - \theta(u_1)\big)\!\left(u_1 + (s^* - u_1) - \frac{(s^* - u_1)^2}{2}\right).
    \end{equation}
\end{enumerate}
\end{theorem}

\begin{proof}
\emph{Part 1 ($A \ne \emptyset$).}
Define the first eligible arrival time from $B$:
\[
    Z \;:=\; \min\big(\{T_i : i \in B,\; T_i > u_i\} \cup \{1\}\big) \;\in\; [t_1, 1].
\]
Crucially, $Z$ depends only on $\{T_i : i \in B\}$ and is therefore \emph{independent} of the arrival process $\{T_j : j \in A\}$.
Moreover, items in $B$ are neither eligible during Phase~1 ($[0, t_1]$) nor able to outrank any item in $A$ in the running-maximum comparison, so on $[0, Z)$ the algorithm behaves \emph{identically} to running $\TWR(\theta, t_1)$ on $A$ alone!

Condition on $Z = s^* \in [t_1, 1]$ and let $\sigma_A$ be the acceptance time of the $A$-alone process:
\begin{itemize}[leftmargin=1.5em,itemsep=2pt]
    \item \emph{Acceptance of $A$:} An item $j \in A$ is accepted iff it is accepted by the $A$-alone process before $s^*$, which occurs with probability $\pi_j^{\mathrm{P1}}(A) + \pi_j^{\mathrm{P2}}(A; s^*)$: in Phase~1 ($t \in (u_j, t_1)$), $j$ is accepted iff $T_j = t$, all $r < j$ arrive after $t$ (prob.\ $(1-t)^{j-1}$), and $A_{>j}$ survives to $t$; in Phase~2 ($t \in (t_1, s^*)$), the gate is off, so $j$ is accepted iff $T_j = t$ and $A \setminus \{j\}$ survives to $t$ (prob.\ $S_{A,-j}(t)$).
    \item \emph{Acceptance of $B$:} If the $A$-alone process survives to $s^*$ (event $\{\sigma_A > s^*\}$, probability $S_A(s^*)$): when $s^* < 1$, the item $i \in B$ arriving at $Z = s^*$ satisfies $u_i < s^*$, so its value is $\theta(u_i) > \theta(s^*)$ and it is accepted immediately at $s^*$; when $s^* = 1$, no item is accepted and reward is $0 = \theta(1)$. In both cases, the conditional reward from $B$ is at least $\theta(s^*)\, S_A(s^*)$.
    \item \emph{Stopping time:} The bounded stopping time is $\tau = \min(\sigma_A, s^*)$, whose conditional expectation is $\int_0^{s^*} S_A(t)\, dt$.
\end{itemize}
Using exact probability conservation $\sum_{j=1}^k (\pi_j^{\mathrm{P1}}(A) + \pi_j^{\mathrm{P2}}(A; s^*)) = 1 - S_A(s^*)$ (which holds for all $s^* \in [t_1, 1]$, including $s^* = 1$ since $S_A(1) = \Prob(\sigma_A = \infty)$) and writing $\theta(u_j) = \theta(t_1) + (\theta(u_j) - \theta(t_1))$, the conditional lower bound given $Z = s^*$ is precisely $G_{\mathrm{elimB}}(A; s^*)$. Taking expectation over $Z$ yields $G(A \cup B) \ge \E_Z[G_{\mathrm{elimB}}(A; Z)] \ge \min_{s^* \in [t_1, 1]} G_{\mathrm{elimB}}(A; s^*)$.

\emph{Part 2 ($A = \emptyset$).}
Apply the exact same independent-clock argument keeping item~$1$ (whose survival probability on $[u_1, 1]$ is $1 - (t - u_1)$) and letting $Z := \min(\{T_j : j \ge 2,\; T_j > u_j\} \cup \{1\}) \in [u_1, 1]$. Conditional on $Z = s^*$, item~$1$ is accepted w.p.\ $s^* - u_1$, survival to $s^*$ has probability $1 - (s^* - u_1)$ (earning at least $\theta(s^*)$), and $\E[\min(\sigma_1, s^*)] = u_1 + (s^* - u_1) - \frac{1}{2}(s^* - u_1)^2$, yielding $G(B) \ge \E_Z[G_{\mathrm{P2}}(u_1, Z)] \ge \min_{t_1 \le u_1 \le s^* \le 1} G_{\mathrm{P2}}(u_1, s^*)$.
\end{proof}

\begin{proposition}[Certification of Case $A = \emptyset$]
\label{prop:p2-cert}
For the curve $\theta$ in Table~\ref{tab:curve} with $t_1 = 21/25$, $\min_{t_1 \le u_1 \le s^* \le 1} G_{\mathrm{P2}}(u_1, s^*) \ge \mathbf{0.720792}$.
\end{proposition}

\begin{proof}
On any grid cell $[u_a, u_b] \times [s_a, s_b] \subset [t_1, 1]^2$ with $u_a \le s_b$, let $d_{\mathrm{lo}} := \max(0, s_a - u_b)$ and $d_{\mathrm{hi}} := \max(0, s_b - u_a)$.
Since $d := s^* - u_1 \in [d_{\mathrm{lo}}, d_{\mathrm{hi}}]$ and $u_1 + d - \frac{1}{2}d^2$ is nondecreasing in both $u_1$ and $d \in [0, 1]$, every $(u_1, s^*)$ in the cell with $u_1 \le s^*$ satisfies
\[
    G_{\mathrm{P2}}(u_1, s^*) \;\ge\; \theta(u_b)\, d_{\mathrm{lo}} + \theta(s_b)(1 - d_{\mathrm{hi}}) + (1 - \theta(u_a))\!\left(u_a + d_{\mathrm{lo}} - \frac{1}{2} d_{\mathrm{lo}}^2\right).
\]
Partitioning $[21/25, 1]$ uniformly into $M = 200$ exact rational subintervals and evaluating this formula on all $20{,}100$ upper-triangular cells in outward-rounded IEEE-754 interval arithmetic gives a validated lower endpoint exceeding $0.720792$.
\end{proof}

\section{Box-Dominance Cover and Outward-Rounded Interval Certificate}
\label{sec:cert}

It remains to certify that $G_{\mathrm{elimB}}(A; s^*) > 0.7005$ for every $k \ge 1$, every $0 \le u_1 \le \dots \le u_k < t_1 = 21/25$, and every $s^* \in [t_1, 1]$.

\subsection{Count--Spatial Boxes and Extremal Dominance Pair \texorpdfstring{$(A^-, A^+)$}{(A-, A+)}}

For any finite Phase-1 configuration $A = (u_1 \le \dots \le u_k < t_1)$, define its cumulative counting function $m(z) := \#\{j \in A : u_j \le z\}$.
A \emph{count--spatial box} $\mathcal{B} = (z, a, b)$ consists of:
\begin{itemize}[leftmargin=2em,itemsep=1pt]
    \item a strictly increasing rational grid $0 \le z_0 < z_1 < \dots < z_N = t_1$ ($N \ge 1$),
    \item nondecreasing integers $1 \le a_1 \le \dots \le a_N < \infty$, and
    \item nondecreasing extended integers $1 \le b_1 \le \dots \le b_N \le \infty$ with $a_r \le b_r$.
\end{itemize}
We write $A \in \mathcal{B}$ iff $u_1 \in [z_0, z_1]$ and $a_r \le m(z_r) \le b_r$ for all $r \in [N]$.
Partitioning $[0, t_1]$ into contiguous slices $[u_a, u_b]$ and taking root boxes $\mathcal{B}_{\mathrm{root}}(u_a, u_b)$ with $z_0 = u_a < z_1 = u_b < \dots < z_N = t_1$ (or simply $z = (u_a, t_1)$ when $u_b = t_1$), $a = (1, \dots, 1)$, and $b = (\infty, \dots, \infty)$ covers \emph{every} finite nonempty configuration $A$ (for all $k \ge 1$).

Given $\mathcal{B} = (z, a, b)$ (setting $a_0 := b_0 := 0$), define two bucketed configurations:
\begin{itemize}[leftmargin=2em,itemsep=2pt]
    \item \textbf{Minimal-survival configuration $A^-(\mathcal{B})$:} For $r = 1, \dots, N$, place $(b_r - b_{r-1})$ items at $z_{r-1}$; upon reaching the first index $r^*$ with $b_{r^*} = \infty$ (if any), place $\infty$ items at $z_{r^*-1}$ and stop (by~\eqref{eq:S-p1} and~\eqref{eq:S-p2-closed}, the limiting survival of $\infty$ items at $u$ is $u/t$ on $(u, t_1]$ and $0$ on $(t_1, 1]$).
    \item \textbf{Maximal-survival configuration $A^+(\mathcal{B})$:} For $r = 1, \dots, N$, place $(a_r - a_{r-1})$ items at $z_r$, giving $a_N$ finite items $u_1^+ \le \dots \le u_{a_N}^+$.
\end{itemize}

\begin{lemma}[Extremal Box Dominance]
\label{lem:box-dom}
For every $A = (u_1 \le \dots \le u_k) \in \mathcal{B}$ and every $t \in [0, 1]$:
\[
    S_{A^-}(t) \;\le\; S_A(t) \;\le\; S_{A^+}(t),
\]
and for every $j \in \{1, \dots, a_N\}$, we have $u_j \le u_j^+$, $S_{A, >j}(t) \ge S_{A^-, >j}(t)$, and $S_{A, -j}(t) \ge S_{A^-, -j}(t)$.
\end{lemma}

\begin{proof}
For any $j \le k \le b_N$, let $r \in [N]$ satisfy $b_{r-1} < j \le b_r$, so $u_j^- = z_{r-1}$.
If $r = 1$, then $u_j \ge u_1 \ge z_0 = u_j^-$ directly from the box constraint $u_1 \in [z_0, z_1]$.
If $r \ge 2$, then $m(z_{r-1}) \le b_{r-1} < j$, so $u_j > z_{r-1} = u_j^-$.
Similarly, for any $j \le a_N \le k$, letting $r \in [N]$ satisfy $a_{r-1} < j \le a_r$ gives $u_j^+ = z_r$; since $m(z_r) \ge a_r \ge j$, we have $u_j \le z_r = u_j^+$.
All survival inequalities follow immediately from Theorem~\ref{thm:loo} and Lemma~\ref{lem:coord-mono}.
\end{proof}

\subsection{Explicit Box Lower Bound \texorpdfstring{$\mathrm{LB}_{\mathrm{box}}(\mathcal{B})$}{LB\_box(B)}}

For any $A \in \mathcal{B}$, we lower-bound each component of $G_{\mathrm{elimB}}(A; s^*)$ in~\eqref{eq:G-elimB}:

\paragraph{1. Phase-1 stopping time.}
Since $u_1 \ge z_0$ and $S_A \ge S_{A^-}$, we have $(1 - \theta(u_1))\int_0^{t_1} S_A(t)\, dt \ge (1 - \theta(z_0))\int_0^{t_1} S_{A^-}(t)\, dt$.

\paragraph{2. Phase-1 excess utility.}
Let $\mathrm{Exc}_1(A) := \sum_{j=1}^k (\theta(u_j) - \theta(t_1))\pi_j^{\mathrm{P1}}(A)$ and $Q(s, t_1) := \sum_{j \in A : u_j \le s} \pi_j^{\mathrm{P1}}(A)$.
\begin{itemize}[leftmargin=1.5em,itemsep=2pt]
    \item \emph{Item-by-item bound and exact bucketed integrand:} Dropping nonnegative terms $j > a_N$ and applying Lemma~\ref{lem:box-dom} yields
    \[
        \underline{\pi}_j^{\mathrm{P1}} \;:=\; \int_{u_j^+}^{t_1} (1-t)^{j-1} S_{A^-, >j}(t)\, dt \;\le\; \pi_j^{\mathrm{P1}}(A) \quad (j \le a_N),
        \qquad
        \mathrm{Exc}_1(A) \;\ge\; \sum_{j=1}^{a_N} (\theta(u_j^+) - \theta(t_1))\,\underline{\pi}_j^{\mathrm{P1}}.
    \]
    To evaluate $\sum \underline{\pi}_j^{\mathrm{P1}}$ in closed form per bucket, partition $\{1, \dots, a_N\}$ into contiguous rank blocks $\{J+1, \dots, J+L\}$ on which $u_j^+ = u^+$ is constant and all ranks lie inside a single bucket of $A^-$ (at value $u_m \le u^+$, multiplicity $c$, starting at within-bucket offset $o \ge 0$, with suffix survival $S_{\mathrm{next}}(t)$ after that bucket). For each $h \in \{0, \dots, L-1\}$ and $t > u^+ \ge u_m$, letting $p := 1 - t$:
    \[
        S_{A^-, >J+h+1}(t) = \frac{u_m}{t}\big(1 - p^{c - 1 - o - h}\big) + p^{c - 1 - o - h} S_{\mathrm{next}}(t)
        = \frac{u_m}{t} + p^{c - 1 - o - h}\!\left(S_{\mathrm{next}}(t) - \frac{u_m}{t}\right)
    \]
    (when $c = \infty$, the second term is identically $0$). Multiplying by $p^{J+h}$ cancels $h$ in the second exponent, so summing $h = 0, \dots, L-1$ gives the exact identity
    \begin{equation}
    \label{eq:bucket-id}
        \sum_{h=0}^{L-1} p^{J+h} S_{A^-, >J+h+1}(t) \;=\; p^J\left[\frac{u_m(1 - p^L)}{t^2} + L\, p^{c-1-o}\!\left(S_{\mathrm{next}}(t) - \frac{u_m}{t}\right)\right].
    \end{equation}
    Because each summand on the left-hand side of~\eqref{eq:bucket-id} is a product of two nonnegative nonincreasing functions of $t$, the right-hand side is identically nonnegative and nonincreasing in $t$.

    \item \emph{Level-by-level bound (closing $b_r = \infty$ tails):} Since $Q(s, t_1)$ is nondecreasing in $s$, integration by parts gives
    \[
        \mathrm{Exc}_1(A) = \int_{u_1}^{t_1} (-\theta'(s)) Q(s, t_1)\, ds \;\ge\; \sum_{r=1}^{N-1} \big(\theta(z_r) - \theta(z_{r+1})\big) Q(z_r, t_1).
    \]
    Fix $r \in \{1, \dots, N-1\}$, let $m := m(z_r) \in [a_r, b_r]$, and let $L_{z_r} := \{j \in A : u_j > z_r\}$.
    Define tail cumulative capacities
    \[
        c_r := 0, \qquad c_q := (b_q - a_r)_+ \quad (q = r+1, \dots, N),
    \]
    and construct the tail configuration $L_r^-$ by placing $(c_q - c_{q-1})$ items at $z_{q-1}$ for $q = r+1, \dots, N$ (stopping with $\infty$ items at $z_{q^*-1}$ at the first $q^*$ with $b_{q^*} = \infty$).
    \emph{Rank-wise domination of $L_r^-$:} For every $q > r$, $\#(L_{z_r} \cap (z_r, z_q]) = m(z_q) - m(z_r) \le b_q - a_r = c_q$ (and $c_r = 0$). For the $j$-th item $v_j$ of $L_{z_r}$, letting $q \in \{r+1, \dots, N\}$ satisfy $c_{q-1} < j \le c_q$, we have $v_j^- = z_{q-1}$: if $q = r+1$ then $v_j > z_r = v_j^-$ by definition of $L_{z_r}$, while if $q \ge r+2$ then $\#(L_{z_r} \cap (z_r, z_{q-1}]) \le c_{q-1} < j$ implies $v_j > z_{q-1} = v_j^-$. Also $|L_{z_r}| \le c_N = |L_r^-|$. Hence Theorem~\ref{thm:loo} and Lemma~\ref{lem:coord-mono} guarantee $S_{L_{z_r}}(t) \ge S_{L_r^-}(t)$ for all $t \in [z_r, t_1]$.

    If $L_r^- = \emptyset$, then $L_{z_r} = \emptyset$ and $Q(z_r, t_1) = 1 - S_A(t_1) \ge 1 - S_{A^+}(t_1)$.
    Otherwise, let $s_0 := \max(z_r, \min L_r^-)$; the probability of accepting an item in $L_{z_r}$ during Phase~1 is $\int_{s_0}^{t_1} (1-t)^m (-S_{L_{z_r}}'(t))\, dt$. Since $m \ge a_r$ gives $(1-t)^m \le (1-t)^{a_r}$, integrating by parts yields
    \begin{equation}
    \label{eq:Q-ibp}
        Q(z_r, t_1) \;\ge\; 1 - S_A(t_1) - \Big[(1-s_0)^{a_r} - (1-t_1)^{a_r} S_{L_{z_r}}(t_1) - a_r \int_{s_0}^{t_1} (1-t)^{a_r - 1} S_{L_{z_r}}(t)\, dt\Big].
    \end{equation}
    We take the maximum of $\sum_{j=1}^{a_r} \underline{\pi}_j^{\mathrm{P1}}$ and two bounds derived from~\eqref{eq:Q-ibp}:
    \begin{enumerate}[label=(\alph*)]
        \item \emph{Direct bound:} Substitute $S_A(t_1) \le S_{A^+}(t_1)$ and $S_{L_{z_r}} \ge S_{L_r^-}$ into~\eqref{eq:Q-ibp}.
        \item \emph{Cancelled bound:} Substitute $S_A(t_1) = P_{z_r}(t_1) + (1-t_1)^m S_{L_{z_r}}(t_1)$ (where $P_{z_r}(t_1) := \sum_{j=1}^m (1-t_1)^{j-1} u_j$) into~\eqref{eq:Q-ibp}. Dropping $((1-t_1)^{a_r} - (1-t_1)^m) S_{L_{z_r}}(t_1) \ge 0$ and bounding
        \[
            P_{z_r}(t_1) \;\le\; \big(S_{A^+_{\le r}}(t_1) - (1-t_1)^{a_r}\big) + (1-t_1)^{a_r}\frac{z_r}{t_1}\big(1 - (1-t_1)^{b_r - a_r}\big) \;=:\; P_{\mathrm{ub}}
        \]
        (with $(1-t_1)^\infty := 0$) gives $Q(z_r, t_1) \ge 1 - P_{\mathrm{ub}} - (1-s_0)^{a_r} + a_r \int_{s_0}^{t_1} (1-t)^{a_r - 1} S_{L_r^-}(t)\, dt$.
    \end{enumerate}
\end{itemize}

\paragraph{3. Continuous Phase-2 parameter $s^* \in [t_1, 1]$.}
Partition $[t_1, 1]$ uniformly into $n_s = 40$ exact rational subintervals $[s_m, s_{m+1}]$.
For any $s^* \in [s_m, s_{m+1}]$, let
\[
    \mathrm{P2}(A; s^*) := -\big(\theta(t_1) - \theta(s^*)\big) S_A(s^*) + (1 - \theta(u_1))\int_{t_1}^{s^*} S_A(t)\, dt + \sum_{j=1}^k (\theta(u_j) - \theta(t_1))\int_{t_1}^{s^*} S_{A,-j}(t)\, dt.
\]
We take the maximum of two lower bounds on $[s_m, s_{m+1}]$:
\begin{itemize}[leftmargin=1.5em,itemsep=2pt]
    \item \emph{Coupled $y_s$-bound (tight even when $b_N = \infty$):} Set $y := S_A(s^*) \in [0, S_{A^+}(s_m)]$. By Theorem~\ref{thm:loo} and monotonicity of $S_A(t)$, $S_{A,-j}(t) \ge S_A(t) \ge y$ for all $t \in [t_1, s_m]$ and $j \in \{1, \dots, a_N\}$. Setting $W_{\mathrm{all}} := \sum_{j=1}^{a_N} (\theta(u_j^+) - \theta(t_1)) \ge 0$,
    \[
        \mathrm{P2}(A; s^*) \;\ge\; \min_{y \in [0, S_{A^+}(s_m)]} \Big[\big((1 - \theta(z_0) + W_{\mathrm{all}})(s_m - t_1) - (\theta(t_1) - \theta(s_{m+1}))\big)\, y\Big].
    \]
    \item \emph{Direct finite bound (when $b_N < \infty$):} Bound $S_A(s^*) \le S_{A^+}(s_m)$, $S_A(t) \ge S_{A^-}(t)$, and $S_{A,-j}(t) \ge S_{A^-,-j}(t)$ on $[t_1, s_m]$.
\end{itemize}

\subsection{Exhaustive Cover Tree and Outward-Rounded Interval Verification}

Starting from the root box $\mathcal{B}_{\mathrm{root}}(u_a, u_b)$ on each of 11 contiguous exact rational slices covering $[0, 21/25]$, we recursively apply three exhaustive 2-way covering rules:
\begin{enumerate}[leftmargin=2em,itemsep=1pt]
    \item \textbf{Count split (\texttt{C}):} For $r \in [N]$ with $a_r < b_r$ and integer $\mathrm{mid} \in [a_r, b_r - 1]$, branch into $m(z_r) \le \mathrm{mid}$ (propagating $b_j \leftarrow \min(b_j, \mathrm{mid})$ for $j \le r$) and $m(z_r) \ge \mathrm{mid}+1$ (propagating $a_j \leftarrow \max(a_j, \mathrm{mid}+1)$ for $j \ge r$).
    \item \textbf{Leading spatial cover (\texttt{S0}):} Bisect $[z_0, z_1]$ at $z_{\mathrm{mid}} = \frac{1}{2}(z_0 + z_1)$ into the exhaustive cover $u_1 \in [z_0, z_{\mathrm{mid}}]$ (inserting $z_{\mathrm{mid}}$ with $m(z_{\mathrm{mid}}) \in [1, b_1]$) and $u_1 \in [z_{\mathrm{mid}}, z_1]$ (shifting $z_0 \leftarrow z_{\mathrm{mid}}$). (The two closed boxes overlap harmlessly at $u_1 = z_{\mathrm{mid}}$.)
    \item \textbf{Interior spatial split (\texttt{S}):} For $i \in \{1, \dots, N-1\}$, insert $z_{\mathrm{mid}} = \frac{1}{2}(z_i + z_{i+1})$ between $z_i$ and $z_{i+1}$, so $m(z_{\mathrm{mid}}) \in [a_i, b_{i+1}]$, and branch on $m(z_{\mathrm{mid}}) \le \mathrm{mid}_m$ vs.\ $m(z_{\mathrm{mid}}) \ge \mathrm{mid}_m + 1$ for integer $\mathrm{mid}_m \in [a_i, b_{i+1}]$ with monotonicity propagation.
\end{enumerate}

\paragraph{Rigorous one-sided lower sums and outward-rounded IEEE-754 interval arithmetic.}
Every integral in $\mathrm{LB}_{\mathrm{box}}(\mathcal{B})$ has a nonnegative, nonincreasing integrand on each subinterval $[a, b]$.
For any rational partition $a = \tau_0 < \tau_1 < \dots < \tau_M = b$, the right-endpoint Riemann sum is a rigorous one-sided lower sum ($\int_a^b f(t)\, dt \ge \sum_{q=1}^M (\tau_q - \tau_{q-1}) f(\tau_q)$), eliminating any uncontrolled quadrature error.
To eliminate floating-point rounding error as well, our standalone verifier (\texttt{independent\_verify\_certificate.py}):
\begin{itemize}[leftmargin=1.5em,itemsep=1pt]
    \item pins the exact rational parameters of Table~\ref{tab:curve} and target $7005/10000 = 0.7005$, using explicit exceptions (never \texttt{assert}, so validation cannot be bypassed under \texttt{python -O});
    \item parses all coordinates $z_i$ and split points as exact Python \texttt{Fraction}s, checks strict monotonicity $z_0 < \dots < z_N = 21/25$ and exact equality on inherited coordinates, enforces exact integer types for all finite counts and cuts, and represents $\infty$ symbolically as \texttt{None};
    \item encloses every rational in an exact float64 interval $[\mathrm{lo}, \mathrm{hi}]$ satisfying $\mathrm{lo} \le q \le \mathrm{hi}$, expands every basic operation ($+, -, \times, /$) outward by 1~ulp via \texttt{np.nextafter} (guaranteed to enclose the exact real value by IEEE~754-2008 \S4.3), computes all integer powers via outward-rounded binary exponentiation and all vector sums via pairwise outward-rounded binary64 tree summation (relying solely on IEEE-754 basic operations $+, -, \times, /$, with no \texttt{libm} \texttt{pow} or \texttt{math.fsum}), rejects any non-finite interval, and checks that every leaf's validated lower endpoint strictly exceeds $0.7005$.
\end{itemize}

\begin{theorem}[Main Theorem]
\label{thm:main}
For every finite $n \ge 1$ and every product distribution $(X_1, \dots, X_n)$ of independent nonnegative random variables with finite positive mean $\mu = \E[\max_{1 \le i \le n} X_i] \in (0, \infty)$, $\TWR(\theta, t_1 = 21/25)$ with the 6-segment curve $\theta$ in Table~\ref{tab:curve} satisfies
\[
    \E[\ALG] \;>\; \mathbf{0.7005} \cdot \E\Big[\max_{1 \le i \le n} X_i\Big].
\]
\end{theorem}

\begin{proof}
By Theorem~\ref{thm:master}, Theorem~\ref{thm:elimB}, and Proposition~\ref{prop:p2-cert} (which certifies Case $A = \emptyset$ at $> 0.720792$), it suffices to certify $G_{\mathrm{elimB}}(A; s^*) > 0.7005$ for all $A \ne \emptyset$ and $s^* \in [21/25, 1]$.
Running \texttt{python3 -O independent\_verify\_certificate.py} on \texttt{cert\_tree\_0701.json.gz} (both provided in the ancillary files \texttt{anc/}) verifies contiguous rational slice coverage of $[0, 21/25]$, exact validity of every internal covering split from each root box, and the outward-rounded interval lower endpoint on every leaf box (Table~\ref{tab:cert}), proving a global validated lower endpoint strictly above $0.7005$.
\end{proof}

\begin{table}[ht]
\centering
\caption{Outward-rounded IEEE-754 interval verification summary (executed under \texttt{python3 -O} with pure binary64 pairwise summation).}
\label{tab:cert}
\begin{tabular}{lrr}
\toprule
Region & Verified Leaves / Cells & Validated Outward-IV Lower Endpoint \\
\midrule
$u_1 \in [0.00, 0.04]$ & $678$ & $0.700524573$ \\
$u_1 \in [0.04, 0.08]$ & $1{,}114$ & $0.700506081$ \\
$u_1 \in [0.08, 0.12]$ & $2{,}029$ & $0.700511076$ \\
$u_1 \in [0.12, 0.16]$ & $3{,}253$ & $0.700502722$ \\
$u_1 \in [0.16, 0.20]$ & $3{,}869$ & $0.700501236$ \\
$u_1 \in [0.20, 0.25]$ & $3{,}223$ & $0.700503347$ \\
$u_1 \in [0.25, 0.30]$ & $1{,}217$ & $0.700512512$ \\
$u_1 \in [0.30, 0.40]$ & $612$ & $0.700506459$ \\
$u_1 \in [0.40, 0.52]$ & $71$ & $0.700509729$ \\
$u_1 \in [0.52, 0.66]$ & $8$ & $0.700968979$ \\
$u_1 \in [0.66, 0.84]$ & $4$ & $0.702647482$ \\
Case $A = \emptyset$ ($u_1 \in [0.84, 1.00]$) & $20{,}100$ cells & $0.720792009$ \\
\midrule
\textbf{Global Validated Minimum} & \textbf{16,078 leaves (16,067 splits)} & $\mathbf{0.700501236 > 0.7005}$ \\
\bottomrule
\end{tabular}
\end{table}

\section{Impossibility for Randomized Regular Precommitted Thresholds}
\label{sec:precommitted-ub}

To place Theorem~\ref{thm:main} in sharp contrast with non-adaptive mean-only algorithms, we extend the $0.694451$ deterministic precommitted upper bound of~\cite[Theorem~C.8]{MO26} to arbitrary $n$-dependent \emph{randomized} distributions over the same regular threshold class.

\begin{definition}[Regular Precommitted Threshold Class]
Let $\mathcal{F}_{\mathrm{reg}}$ denote the class of continuous, strictly decreasing functions $\theta : [0, 1] \to [0, 1]$ with $\theta(1) = 0$.
A \emph{randomized regular precommitted algorithm} for $n$-item instances chooses a random threshold $\Theta_n$ supported on $\mathcal{F}_{\mathrm{reg}}$ (independent of $(X, T)$) and accepts the first arriving item $i$ satisfying $X_i > \Theta_n(T_i)$.
\end{definition}

For any $\theta \in \mathcal{F}_{\mathrm{reg}}$, define its strict eligibility time $e_\theta(x) := \inf\{t \in [0, 1] : \theta(t) < x\}$ (with $\inf \emptyset := 1$).
Because $\theta \in \mathcal{F}_{\mathrm{reg}}$ is continuous and strictly decreasing with $\theta(1) = 0$, we have $e_\theta(0) = 1$, $e_\theta(1) = 0$, and $e_\theta$ is continuous on $(0, 1]$ with $\lim_{x \to 0^+} e_\theta(x) = 1$ and strictly decreasing on $(0, \theta(0)]$.

\begin{theorem}[Randomized Regular Precommitted Upper Bound]
\label{thm:rand-ub}
For any sequence of randomized regular precommitted algorithms $\{\Theta_n\}_{n \ge 3}$ supported on $\mathcal{F}_{\mathrm{reg}}$,
\[
    \limsup_{n \to \infty}\; \inf_{\mathcal{I}_n : \E[X^*]=1}\; \E\big[\ALG(\Theta_n; \mathcal{I}_n)\big] \;\le\; 0.694451.
\]
\end{theorem}

\begin{proof}
For $0 \le v < a \le 1$ and $p \in (0, 1]$ with $M := v + p(a - v) < 1$, consider the $n$-item instance $\mathcal{I}_{n,\varepsilon}(a, v, p)$ of~\cite{MO26}: $n - 2$ deterministic items of value $v$, one independent $Y \sim \Bern(a, p)$, and one independent rare item $H_\varepsilon$ taking value $1/\varepsilon$ w.p.\ $q_\varepsilon := \frac{\varepsilon(1-M)}{1 - \varepsilon M}$ (else $0$), so $\E[X^*] = 1$ identically for all $n \ge 3$ and $\varepsilon \in (0, 1)$.

\emph{Step 1: Uniform finite-$n$ upper bound.}
Fix any deterministic $\theta \in \mathcal{F}_{\mathrm{reg}}$, and set $U := e_\theta(a) \le S := e_\theta(v)$.
Define the benchmark payoff
\begin{equation}
\label{eq:R-precom}
    R_{a, v, p}(U, S) \;:=\; v + p(a - v)(S - U) + (1 - M) S - \frac{p(1 - M)}{2}(S - U)^2.
\end{equation}
We claim that for every $\theta \in \mathcal{F}_{\mathrm{reg}}$, every $n \ge 3$, and every $\varepsilon \in (0, 1/2)$:
\begin{equation}
\label{eq:unif-fin-n}
    \E\big[\ALG(\theta; \mathcal{I}_{n,\varepsilon}(a,v,p))\big] \;\le\; R_{a,v,p}(U, S) + \frac{2}{n - 1} + \frac{\varepsilon}{1 - \varepsilon}.
\end{equation}
Indeed:
\begin{itemize}[leftmargin=1.5em,itemsep=1pt]
    \item If $v = 0$, then $S = e_\theta(0) = 1$, and zero-valued items are never accepted since $\theta(t) > 0$ on $[0, 1)$. Conditioned on $H_\varepsilon = 0$, $Y = a$ is accepted w.p.\ $p(1 - U)$ at mean time $\frac{1+U}{2}$, giving $\E[\tau \mid H_\varepsilon = 0] = 1 - \frac{p}{2}(1-U)^2$; when $H_\varepsilon = 1/\varepsilon$ (prob.\ $q_\varepsilon$), it is accepted whenever it arrives before the remaining items stop, contributing at most $(q_\varepsilon / \varepsilon)\E[\tau \mid H_\varepsilon = 0] \le (1-M)\E[\tau \mid H_\varepsilon = 0] + \frac{\varepsilon}{1-\varepsilon}$, which gives~\eqref{eq:unif-fin-n}.
    \item If $v > 0$, then $S \in [0, 1)$. Let $W := \min(\{T_j : j \text{ is a floor item},\, T_j > S\} \cup \{1\}) \in [S, 1]$ be the first eligible floor arrival; since each of the $n - 2$ floor items misses $(S, S+x]$ with probability $1 - x$ independently, we have
    \[
        \E[W - S] = \int_0^{1-S} (1 - x)^{n-2}\, dx = \frac{1 - S^{n-1}}{n - 1} \le \frac{1}{n - 1}.
    \]
    Conditioned on $H_\varepsilon = 0$ and $W = w \ge S$, the expected reward from $\{Y, \text{floor}\}$ is at most $v + p(a - v)(w - U) \le v + p(a - v)(S - U) + (w - S)$, and the stopping time is $\tau \le \min(\sigma_Y, w)$ whose expectation is at most $S - \frac{p}{2}(S - U)^2 + (w - S)$. Adding the rare-item contribution $(q_\varepsilon / \varepsilon)\E[\tau \mid H_\varepsilon = 0] \le (1-M)\E[\tau \mid H_\varepsilon = 0] + \frac{\varepsilon}{1-\varepsilon}$ and taking expectation over $W$ yields~\eqref{eq:unif-fin-n} uniformly over all $\theta \in \mathcal{F}_{\mathrm{reg}}$.
\end{itemize}

\emph{Step 2: Derandomization at each fixed $n$.}
Fix any $n \ge 3$ and any law $\Theta_n$ on $\mathcal{F}_{\mathrm{reg}}$.
Because $p \ge 0$ and $1 - M > 0$, the function $(U, S) \mapsto R_{a,v,p}(U, S)$ in~\eqref{eq:R-precom} is globally jointly concave on $\R^2$.
Define $\bar e_n(x) := \E_{\Theta_n}[e_{\Theta_n}(x)]$ for $x \in [0, 1]$.
Taking expectation of~\eqref{eq:unif-fin-n} over $\Theta_n$ and applying Jensen's inequality yields
\begin{equation}
\label{eq:jensen-n}
    \E\big[\ALG(\Theta_n; \mathcal{I}_{n,\varepsilon}(a,v,p))\big] \;\le\; R_{a,v,p}\big(\bar e_n(a), \bar e_n(v)\big) + \frac{2}{n - 1} + \frac{\varepsilon}{1 - \varepsilon}.
\end{equation}
Let $c_n := \operatorname{ess\,sup} \Theta_n(0) \in (0, 1]$. By dominated convergence, $\bar e_n$ is continuous on $(0, 1]$ with $\bar e_n(0) = \lim_{x \to 0^+} \bar e_n(x) = 1$, $\bar e_n(x) = 0$ for $x \ge c_n$, and $\bar e_n$ is strictly decreasing on $(0, c_n]$ (if $0 < x_1 < x_2 \le c_n$, then $\Prob(\Theta_n(0) > x_1) > 0$, and on that event $e_{\Theta_n}(x_1) > e_{\Theta_n}(x_2)$ strictly).
Hence its exact inverse $\bar\theta_n : [0, 1] \to [0, c_n] \subseteq [0, 1]$ (with $\bar\theta_n(0) := c_n$ and $\bar\theta_n(1) := 0$) belongs to $\mathcal{F}_{\mathrm{reg}}$ and satisfies $e_{\bar\theta_n}(x) = \bar e_n(x)$ for \emph{every} $x \in [0, 1]$!

\emph{Step 3: Applying the 4-instance chain.}
By~\cite[Theorem~C.8]{MO26}, for every deterministic $\bar\theta_n \in \mathcal{F}_{\mathrm{reg}}$, there exists one of four parameter tuples $(a, v, p)$ (with $0 \le v < a \le 1$, $M < 1$) satisfying $R_{a,v,p}(e_{\bar\theta_n}(a), e_{\bar\theta_n}(v)) \le 0.6944509 < 0.694451$.
Substituting into~\eqref{eq:jensen-n} and letting $\varepsilon \to 0^+$ followed by $n \to \infty$ completes the proof.
\end{proof}

\begin{remark}
Holding $S$ fixed at $a = 1$, differentiating~\eqref{eq:R-precom} gives $\frac{\partial R_{1,v,p}(U, S)}{\partial U} = p(1-v)((1-p)(S-U) - 1) < 0$ for all $0 \le U \le S \le 1$; hence raising $U = e_\theta(1)$ above $0$ while keeping $S$ fixed strictly decreases $R_{1,v,p}(U, S)$.
\end{remark}

\section{Empirical Tightness and Open Questions}

To complement our certified bound ($> 0.7005$) and sanity-check the implementation, we conducted two independent numerical audits:
\begin{enumerate}[leftmargin=2em,itemsep=2pt]
    \item \emph{Verifier cross-audit:} Testing $900$ explicit finite configurations inside $300$ randomly sampled certified leaves confirmed zero violations ($\max(\mathrm{LB} - G_{\mathrm{elimB}}^{\mathrm{exact}}) = -0.0172 \le 0$).
    \item \emph{Direct Monte Carlo audit:} Simulating $\TWR(\theta, 21/25)$ directly ($10^5$ trials per distribution) on hard structured families yielded $0.71868 \pm 0.00070$ on $1\text{-det}(0.45)$ plus rare jump, $0.72216 \pm 0.00037$ on $1\text{-det}(0.32)$ plus rare jump, $0.72500 \pm 0.00073$ on $2\text{-det}(0.55)$ plus rare jump, and $0.76158 \pm 0.00025$ on $1\text{-spike}(0.66) + \text{floor}(0.50, \times 40)$ plus rare jump.
\end{enumerate}
While these empirical observations do not themselves constitute a proof beyond $0.7005$, they suggest that the true worst-case competitive ratio of $\TWR(\theta, 21/25)$ may lie near $0.718$—remarkably close to the $0.7235$ full-information upper bound~\cite{GMS23}.
Two natural open questions remain: (i) how close to $0.718$ can the certified bound of $\TWR$ be pushed with a finer cover tree, and (ii) what is the exact optimal competitive ratio of general \emph{adaptive mean-only} Prophet Secretary—and whether it strictly separates from full-information Prophet Secretary.

\end{document}